\documentclass[usenatbib]{mn2e}
\usepackage{times}
\usepackage{graphicx}
\usepackage{graphics} 
\usepackage{epsfig}       

\makeatletter
\renewcommand*{\p@section}{\S\,}
\renewcommand*{\p@subsection}{\S\,}
\makeatother
         
\def\aj{{AJ}}                   
\def\araa{{ARA\&A}}          
\def\apj{{ApJ}}                 
\def\apjl{{ApJ}}                
\def\apjs{ {ApJS}}

\def\aap{ {A\&A}}

\def\mnras{ {MNRAS}}

\def\nat{ {Nature}}

\newcommand{\be}{\begin{equation}}
\newcommand{\ee}{\end{equation}}

\newcommand{\msun}{{$M_{\odot}$}}

\newcommand{\es}{erg~s$^{-1}$}

\newcommand{\esc}{erg~s$^{-1}$~cm$^{-2}$}
\newcommand{\degree}{$^{\circ}$} 
\newcommand{\gtsima}{$\; \buildrel > \over \sim \;$}
\newcommand{\ltsima}{$\; \buildrel < \over \sim \;$}
\newcommand{\prosima}{$\; \buildrel \propto \over \sim \;$}
\newcommand{\gsim}{\lower.5ex\hbox{\gtsima}}
\newcommand{\lsim}{\lower.5ex\hbox{\ltsima}}
\newcommand{\simgt}{\lower.5ex\hbox{\gtsima}}
\newcommand{\simlt}{\lower.5ex\hbox{\ltsima}}
\newcommand{\simpr}{\lower.5ex\hbox{\prosima}}

\newcommand{\etal}{{et al.}}
\newcommand{\cxo}{\textit{Chandra}}

\newcommand{\lx}{$L_{\rm X}$}
\newcommand{\lr}{$L_{\rm r}$}

\newcommand{\ledd}{$L_{\rm Edd}$}

\newcommand{\ellr}{$\ell_{\rm r}$}
\newcommand{\ellx}{$\ell_{\rm X}$}

\newcommand{\ratio}{$L_{\rm X}$/$L_{\rm Edd}$}
\newcommand{\so}{XTE~J1118+480}
\newcommand{\gx}{GX339-4}
\begin{document}

\title[BH X-ray binary radio/X-ray domain]{The radio/X-ray domain of black hole X-ray binaries at the lowest radio luminosities}
\author[E. Gallo \etal]{E. Gallo$^{1}$\thanks{egallo@umich.edu}, J. C. A. Miller-Jones$^{2}$, D. M. Russell$^3$, 
P. G. Jonker$^{4,5,6}$, J. Homan$^{7}$ \and R. M. Plotkin$^{1}$, S. Markoff$^{8}$, B. P. Miller$^{9}$, S. Corbel$^{10}$ and R. P. Fender$^{11}$\smallskip\\
$^{1}${Department of Astronomy, University of Michigan, 500 Church St., Ann Arbor, MI 48109, USA}\\
$^{2}${International Centre for Radio Astronomy Research, Curtin University, GPO Box U1987, Perth, WA 6845, Australia}\\
$^{3}${New York University Abu Dhabi, P.O. Box 129188, Abu Dhabi, United Arab Emirates}\\
$^{4}${SRON, Netherlands Institute for Space Research, Sorbonnelaan 2, 3584~CA, Utrecht, The Netherlands}\\
$^{5}${Harvard--Smithsonian Center for Astrophysics, 60 Garden Street, Cambridge, MA~02138,USA}\\
$^{6}${Department of Astrophysics/IMAPP, Radboud University Nijmegen, 6500 GL, Nijmegen, The Netherlands}\\
$^{7}${MIT Kavli Institute for Astrophysics and Space Research, 70 Vassar Street, Cambridge, MA 02139, USA}\\
$^{8}${Astronomical Institute `Anton Pannekoek', University of Amsterdam, 1090 GE Amsterdam, The Netherlands}\\
$^{9}${Physics and Astronomy Department, Macalester College, Saint Paul, MN 55105, USA}\\
$^{10}${AIM-Unit\'e Mixte de Recherche CEA-CNRS, UniversitŽ Paris VII, CEA Saclay, Service d'Astrophysique, F-91191 Gif-sur-Yvette, France}\\
$^{11}${Astrophysics, Department of Physics, University of Oxford, Keble Road, OX1 3RH, Oxford, United Kingdom}}
\maketitle
\begin{abstract}
We report on deep, coordinated radio and X-ray observations of the black hole X-ray binary \so\ in quiescence. The source was observed with the Karl G. Jansky Very Large Array for a total of 17.5 hrs at 5.3 GHz, yielding a  $4.8\pm 1.4$ $\mu$Jy radio source at a position consistent with the binary system. At a distance of 1.7 kpc, this corresponds to an integrated radio luminosity between 4-8 $\times 10^{25}$ \es, depending on the spectral index. This is the lowest radio luminosity measured for any accreting black hole to date. Simultaneous observations with the \textit{Chandra X-ray Telescope} detected \so\ at $1.2\times 10^{-14}$ \esc\ (1-10 keV), corresponding to an Eddington ratio of $\sim$$4\times 10^{-9}$ for a 7.5 \msun\ black hole. Combining these new measurements with data from the 2005 and 2000 outbursts available in the literature, we find evidence for a relationship of the form \ellr=$\alpha$$+$$\beta$\ellx\ (where $\ell$ denotes logarithmic luminosities), with $\beta=0.72\pm0.09$. \so\ is thus the third system -- together with \gx\ and V404 Cyg -- for which a tight, non-linear radio/X-ray correlation has been reported over more than 5 dex in \ellx. 
Confirming previous results, we find no evidence for a dependence of the correlation normalisation of an individual system on orbital parameters, relativistic boosting, reported black hole spin and/or black hole mass.
We then perform a clustering and linear regression analysis on what is arguably the most up-to-date collection of coordinated radio and X-ray luminosity measurements from quiescent and hard state black hole X-ray binaries, including 24 systems.  At variance with previous results, a two-cluster description is statistically preferred only for random errors $\simlt$0.3 dex in both \ellr\ and \ellx, a level which we argue can be easily reached when the known spectral shape/distance uncertainties and intrinsic variability are accounted for. A linear regression analysis performed on the whole data set returns a best-fitting slope $\beta=0.61\pm0.03$ and intrinsic scatter $\sigma_0=0.31\pm 0.03$ dex.
\end{abstract}
%
%
\section{Introduction}
Coordinated radio and X-ray monitoring of black hole X-ray binaries (BHBs) has established as a powerful diagnostics for the connection between accretion and the production of relativistic jets in these systems. A tight and repeating non-linear correlation between the radio and X-ray luminosity, of the form \ellr=$\alpha$$+$$\beta$\ellx (where $\ell$ denotes logarithmic values), with $\beta\simeq 0.7$, was first established for the BHB GX339-4 in the hard state by \citeauthor{corbel03} (2003, see also \citealt{corbel00, hannikainen98}), and later confirmed with data from 7 outbursts over a period of 15 yr \citep{corbel13}.
Shortly afterward, \cite{gfp03} reported that the same relation was exhibited by the BHB V404 Cyg (see \citealt{corbel08} for an update), and that quasi-simultaneous radio and X-ray luminosity measurements for 8 more hard state systems (albeit each spanning quite a limited dynamic range compared to \gx\ and V404 Cyg) were consistent with the same scaling, arguing for a `universal' scaling relation holding all the way from X-ray Eddington ratios (\ratio) as low as $10^{-5}$ up to $10^{-2}$, above which the radio emission from the steady jet is suppressed \citep{fender99, russell11}.   

The non-linearity of the correlation is consistent with scale-invariant jet models where a flat-spectrum, partially self-absorbed jet is coupled to radiatively inefficient accretion, yielding a scaling of the X-ray luminosity with the second power of the accretion rate \citep{falcke95, heinz03, markoff03}. The simultaneous radio/X-ray detection of A0620-00 at \ratio$\simeq 10^{-8.5}$ \citep{gallo06} seemed to confirm that the correlation extends down to very low quiescent luminosities, with no obvious break (such as predicted by \citealt{yuancui}). 

Since, the universality of the radio/X-ray correlation for BHBs has been called into question. New observations of known sources, along with newly discovered ones, have resulted in an increasingly large number of outliers lying well outside the scatter about the previously established best-fitting relation \citep{corbel04, brocksopp05, cadolle07, rodriguez07, gallo07, xuecui, jonker10, coriat11, solerifender, soria11, ratti12, brocksopp13}.  In \citeauthor{gallo12} (2012, GMF12 hereafter), making use of state of the art data clustering algorithms, we showed that a dual cluster model, with independent linear fits (in log space), was a significant improvement over fitting all points from 18 BHBs as a single cluster.  The fact that the lower track appears to populate the high-luminosity end of the radio/X-ray plane is not necessarily dictated by sensitivity limitations, as there is tentative evidence for as many as 3 sources {jumping from the lower to the higher track as they fade towards quiescence} (i.e., H 1743$-$322, MAXI~J1659$-$152 and Swift J1357.2$-$0933; see respectively \citealt{jonker10, jonker12} and Kolehmainen \etal, in prep.). This behaviour has been interpreted as indicative of  radiatively efficient accretion taking place in the hard state in some sources during the first phases of an outburst decline \citep{coriat11}, possibly due to the transient emergence of a cool accretion disc in the very innermost region of an otherwise inefficient accretion flow \citep{m-hm14}. 

With A0620--00 being the only anchor of the upper track in the truly quiescent regime (i.e. \ratio$\simeq 10^{-8.5}$), however, additional constraints from quiescent sources are highly desirable to further probe the demographics of the radio/X-ray luminosity plane BHBs, and its physical interpretation. With this goal in mind, \cite{miller-jones11a} set out to carry out a deep radio survey of low X-ray luminosity systems with the upgraded Karl G. Jansky Very Large Array (VLA; see also \citealt{calvelo10}).  
Out of a handful of truly quiescent sources that were targeted -- with distances ranging from 1.8 to 9.6 kpc -- only one yielded a $>2 \sigma$ radio detection. {While not formally significant}, the measured radio brightness at the known source position of XTE J1118+480 was $6.4\pm2.6~\mu$Jy/beam.
This system, hosting a dynamically confirmed black hole accretor at a distance of 1.7$\pm$0.1 kpc (\citealt{gelino06}; however, see \ref{sec:disc} for caveats related to the quoted distance uncertainty), has been in quiescence since the end of its 2005 outburst \citep{zurita06}. Sensitive X-ray observations in quiescence, with the \textit{Chandra X-ray Telescope}, date back to 2002 January, about 2 years after the decay from its previous outburst, in 2000 \citep{chaty03}. In 2002, the source was detected at \lx $\simeq 3.5 \times  10^{30}$ \es~or $\simeq 10^{-8.5}$ times its Eddington luminosity (for a black hole mass between 6.9 and 8.2 \msun; \citealt{khargharia13}). The composite optical-UV-X-ray spectrum, obtained via simultaneous Multiple Mirror Telescope and \textit{Hubble Space Telescope} observations, was interpreted in the context of an advection dominated accretion flow model, with a transition radius of about $10^4$ Schwarzschild radii \citep{mcclintock03}. However, the lack of coordinated radio/IR coverage represented a severe limitation to the interpretation of the system's Spectral Energy Distribution (SED). 

In this paper, we report on new, deep, coordinated \cxo~ and VLA observations of XTE~J1118+480, i.e. one of less than a handful of known sources for which we could realistically hope to either secure a simultaneous radio/X-ray detection or make use of deeper upper limits to constrain the slope of the empirical radio/X-ray relation down to the lowest measurable luminosities.  
In a companion paper (Plotkin \etal, submitted), we report on optical and infrared observations simultaneous to the VLA and \cxo\ observations presented here, and discuss a possible interpretation of the system's broadband SED in quiescence in the contest of radiative jet model, along with its implications for constraining the jet acceleration across different regimes. 
%
%
\begin{figure}
\psfig{figure=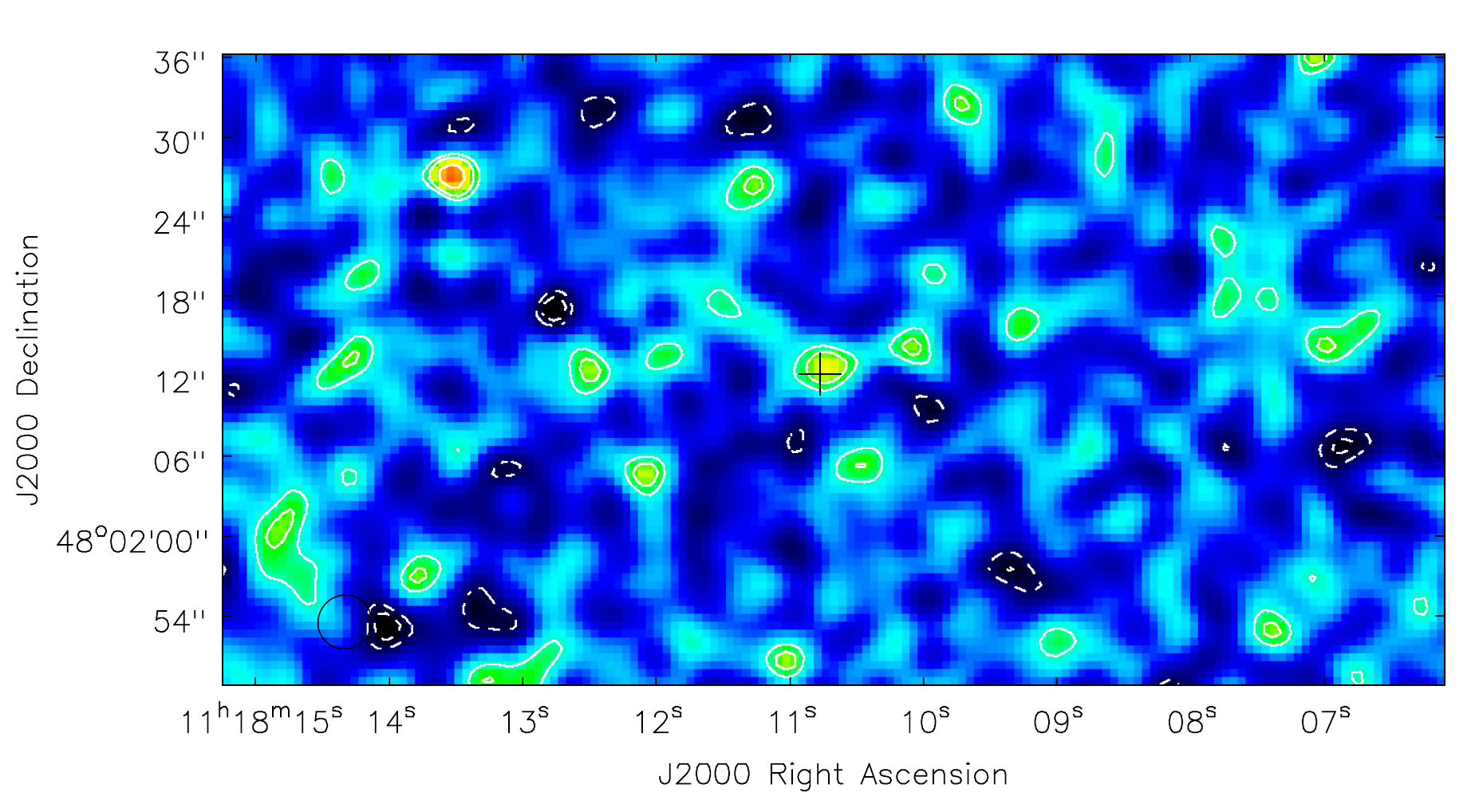,width=0.485\textwidth,angle=0}
\caption{Karl G. Jansky VLA image of the field of view of \so. The colour scale runs from 4 to 10 $\mu$Jy beam$^{-1}$, and the contours at $\pm 2 \times (\sqrt{2})^n$ times the rms noise level of 1.36 $\mu$Jy beam$^{-1}$. The known Very-Long-Baseline Interferometry (VLBI) position of the BHB system \citep{mirabel01} is marked by a cross.}\label{fig:vla}
\end{figure}
 %
 %
\section{Observations and data reduction}
%
%
\subsection{Karl G. Jansky Very Large Array}
In order to confirm the marginal-significance detection of \citet{miller-jones11a}, we required a significantly lower noise level. 
We observed XTE J1118+480 with the Karl G. Jansky VLA on 2013 June 27 and 28 (MJD 56471 and 56472), for 7.5\,h on each day (0830-1600 LST).  We observed in full polarization mode with two overlapping 1024-MHz basebands, centered at frequencies of 4.8 and 5.8\,GHz.  Each baseband was composed of eight contiguous 128-MHz sub-bands, each comprising sixty-four 2\,MHz spectral channels.  The array was in the relatively compact C configuration, giving an angular resolution of $\sim4$\,arcsec.  A power outage at the VLA site caused us to lose 2\,h of observation on June 28th, leaving us a total of 11.3\,h of time on source.  We offset the pointing position by 7\,arcsec (2 synthesised beams) from the known VLBI position of XTE J1118+480 \citep{mirabel01}, to prevent artifacts generated at the phase center by correlation errors from creating a spurious source.

3C\,286 was used as both bandpass and amplitude calibrator, and the secondary calibrator was J1126+4516, a 0.4-Jy source located 3.2$^{\circ}$ away from the target, XTE J1118+480.  We reduced and imaged the data with the Common Astronomy Software Application (\textsc{CASA}) v4.1.0, using standard procedures.  The data were Hanning smoothed, and then edited to remove radio frequency interference.  
Bandpass calibration was carried out before the amplitude and phase gains were derived for both calibrator sources, using the `Perley-Butler 2010' coefficients within the CASA task SETJY to set the amplitude scale \citep{perley13}.  The complex gain solutions derived for the secondary calibrator were interpolated to the target source, before averaging the resulting calibrated data by a factor of three in frequency to reduce the data volume.
Imaging and self-calibration were then performed separately for each day's observations.  The data were imaged out to 15\,arcmin, well beyond the distance to the half-power point of the primary beam.  We used Briggs weighting (robust=1) as a compromise between sensitivity and suppression of side-lobes from bright sources elsewhere in the field.  We used the $w$-projection algorithm to prevent phase errors due to sky curvature from affecting our deconvolution, and modelled the sky frequency dependence using two Taylor terms.  The brightest confusing source in the field was NVSS J111820+475659, with a flux density (without primary beam correction) of 4.3\,mJy\,beam$^{-1}$.  There was sufficient emission in the field to perform self-calibration, initially solving only for phases, and then for amplitude and phase, down to solution intervals of 1 and 5\,minutes, respectively.  For the shorter solution intervals, data from all spectral windows were combined prior to solving for the time-dependent gains, to provide sufficient signal-to-noise to give robust solutions. %
%
\begin{figure}
\psfig{figure=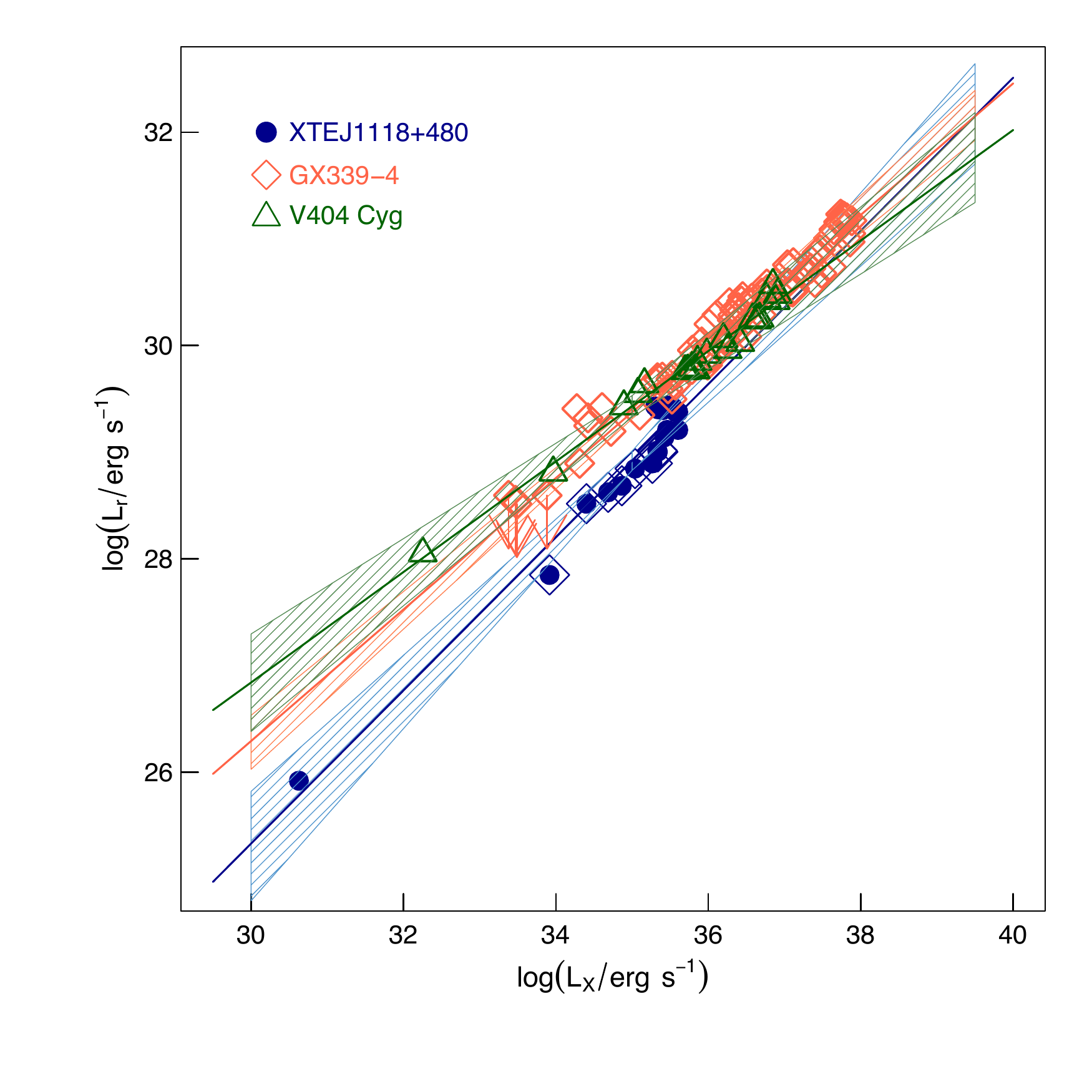,width=0.5\textwidth,angle=0}
\caption{Linear regression analysis for V404 Cyg (open green triangles; data from \citealt{corbel08}), \gx\ (open orange diamonds; data from  \citealt{corbel13}) and \so\ (filled blue circles). For the latter, data are taken from this work (lowest \lx\ and \lr\ luminosity point) plus Hynes \etal\ (2000) and Fender \etal\ (2001) for the 2000 outburst, and Brocksopp \etal\ (2010) plus Dunn \etal\ (2010) for the 2005 outburst, which is highlighted by open blue diamond symbols encircling the filled blue circles. 
See Table 1 for a complete list of the best-fitting parameters. } \label{fig:3sources}
\end{figure}
Finally, we combined the two self-calibrated data sets to provide the deepest possible image (shown in Figure \ref{fig:vla}), reaching an rms noise level of 1.45\,$\mu$Jy\,beam$^{-1}$.  A $3.2\sigma$ peak was detected 0.59\,arcsec from the predicted source position (taking into account the expected proper motion measured by \citealt{mirabel01}), well within the beamsize of $4.2\times4.0$\,arcsec$^2$.  To improve the significance of the detection, we then combined our new data from 2013 June with archival data from 2010 November, providing an extra 2.4\,h of time on source.  This reduced the noise level to 1.36\,$\mu$Jy\,beam$^{-1}$ in the region around the target.  Fitting the emission at the target position with a point source in the image plane gave a flux density of $4.79\pm1.45$\,$\mu$Jy\,beam$^{-1}$, where the quoted uncertainty represents the rms noise added in quadrature to the uncertainty on the point source fit.
Approximating the integrated radio luminosity as the monochromatic luminosity multiplied by the observing frequency, this corresponds to a radio luminosity \lr $=8.3\times 10^{25}$ \es. This assumes a flat radio spectrum and a minimum synchrotron emitting frequency much smaller than the observing frequency, though the former assumption is violated by some hard state black hole X-ray binaries; e.g., if a spectral index $a=+0.5$ is assumed (where the flux density scales as $S_{\nu} \propto \nu^{+a}$), the resulting integrated radio luminosity is factor about 2 lower.
%
%
\subsection{\textit{Chandra X-ray Telescope}}
\so\ was observed with \cxo\ ACIS-S on 2013 June 27 (PI: Gallo; ObsId 14630); the data were telemetered in Very Faint (VF) mode, with a high-energy cutoff at 13 keV, and analyzed with the Chandra Interactive Analysis of Observations (\textsc{CIAO}) software, v4.5. 
Event files were reprocessed with the \textsc{CIAO} script {\textsc{chandra$\_$repro}}. No flares were detected in the background light curve, yielding a net exposure time of 58 ks. The analysis described below was carried out between 0.5-7 keV, where the instrument is best calibrated. 

An X-ray source was clearly detected at a position consistent with \so; in order to estimate bounds to the source net count rate and energy flux, we made use of the newly-developed CIAO script {\textsc{aprates}}, which employs Bayesian statistics to compute the posterior probability distribution for the source intensity assuming non-informative priors. 
Source and background parameters were extracted from a circular region with a 2\,arcsec radius and an annulus of inner and outer radius of 10 and 30\,arcsec respectively, both centered at the X-ray source peak position. 

First, we simulated the source and background Point Spread Function (PSF) with the Chandra Ray Tracer (\textsc{ChaRT}), using the source and background spectra as extracted from the \textsc{specextract} script as an input. The PSF event files were then created by projecting the ChaRT output onto the detector plane using the \textsc{Marx} software (v4.5). The observation exposure map was generated with the \textsc{flux\_image} CIAO script. To estimate net counts, \textsc{aprates} requires the following inputs: the source/background region counts and areas (134/358 counts; 116.8/10382.7 pixels); the PSF fraction in the source aperture (0.953) and the PSF fraction in the background aperture (0.029). These inputs yield  a net count rate of 2.26$\times 10^{-3}$ cps, with the 90 per cent confidence region extending from 1.93-2.59$\times 10^{-3}$cps. The net energy flux can be estimated using the exposure map to set the average photon energies in the source and background apertures; this gives a 0.5-7 keV absorbed flux of 1.44$\times 10^{-14}$ \esc, with  90 per cent confidence bounds between 1.22 and 1.66$\times 10^{-14}$ \esc.
Assuming the standard value of 1.3$\times 10^{20}$ cm$^{-2}$ for the Galactic hydrogen equivalent column towards \so\ \citep{mcclintock01a, reis09}, and a power law spectrum with photon index $\Gamma=2.1$ (typical of quiescent black hole X-ray binaries; \citealt{plotkin13}), this corresponds to an unabsorbed 1-10 keV  (to allow for a proper comparison with \citealt{corbel13}) flux of 1.2$\times 10^{-14}$ \esc, that is, a luminosity of 4.3$\times 10^{30}$ \es\ for a distance of 1.7 kpc.
%
%
\begin{figure*}
\psfig{figure=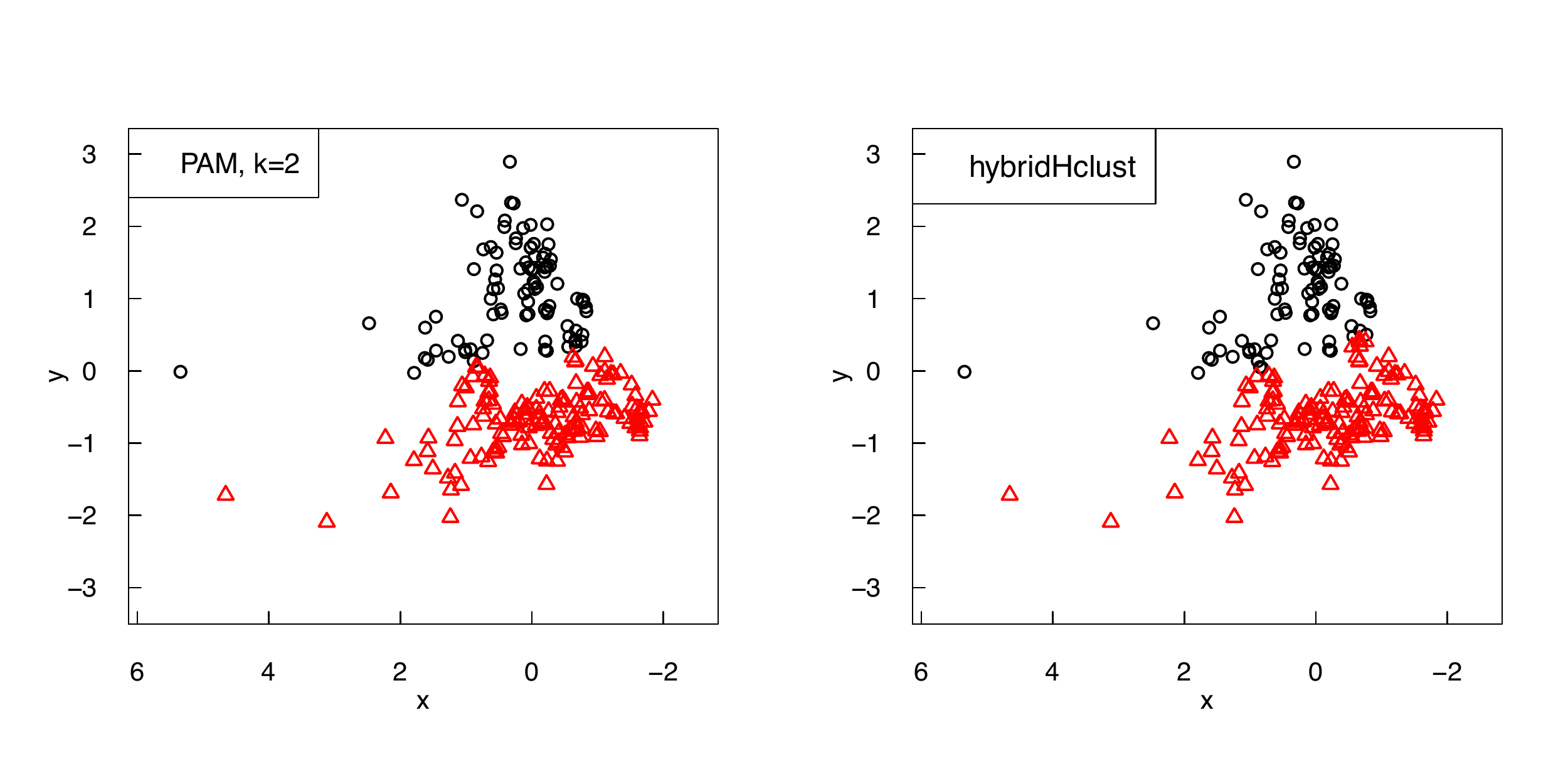,width=0.85\textwidth,angle=0}
\caption{Shown are nearly-simultaneous radio and X-ray luminosity measurements for 24 hard and quiescent state black hole X-ray binaries, taken from GMF12 plus Corbel \etal (2013). Standardized $x$:$y$ coordinates are obtained starting from initial radio and X-ray luminosity data set by $i)$ taking the logarithm; $ii)$ subtracting the mean, and $iii)$ dividing by the standard deviation. The major component vector is then identified by applying principal component analysis to the resulting vectors. The new coordinates are finally rotated in order to align with the principal component, and scaled to unit variance. After running the principal component analysis, a formal clustering analysis is performed on the standardized coordinates. Left: Results from the `partitioning around medoids'  method (\textsc{PAM}; \citealt{pam}). Right: Results from the `hybrid hierarchical clustering via mutual clusters' method (\textsc{hybridHclust}; \citealt{hybridclust}). Although both methods identify two clusters, the cluster memberships are not entirely consistent with each other; specifically, different splits are returned around ($x$=1:$y$=0) and ($x$=$-$0.7:$y$=0).}\label{fig:clust}
\end{figure*}
%
%
\begin{table}
\caption{Best fitting parameters for the three sources in Figures \ref{fig:3sources} and \ref{fig:finale}. Linear regression analysis has been performed to test a relation of the form (\ellr $- \ell_{\rm r,c}$) $= \alpha +\beta($\ellx$ -\ell_{\rm X,c}$). Uncorrelated random errors of 0.3 dex are adopted in all fits. Centring values, $\ell_{\rm X,c}$ and $\ell_{\rm r,c}$, are fixed to the median values for the whole sample (24 systems), i.e., 35 and 29, respectively. Columns are: (1) Best-fitting intercept, $\alpha$; (2)  Best-fitting slope, $\beta$; (3) Best-fitting intrinsic scatter, $\sigma_0$.  Quoted 1$\sigma$ errors correspond to the 16$^{\rm th}$ and $84^{\rm th}$ percentiles. 
\label{table:fits}}
\begin{tabular}{lllll}
\hline
				&	GX339-4 	 		& 	V404~Cyg 			&	J1118$+$480 	& 		24 BHBs				\\
				& 	(a)					&	(b)					&		(c)		&    					  		\\
\hline	
$\alpha$ 			&	$0.37 \pm 0.06$		&	$0.43 \pm 0.10$	&	$-0.08\pm 0.09$		&	$0.15\pm 0.03$	\\
$\beta$ 			&	$0.62 \pm 0.04$		& 	$0.52 \pm 0.07$ 	&	$0.72\pm 0.09$		& 	$0.61\pm 0.03$		\\
$\sigma_0$ 		&	$0.05 \pm 0.03 $		&	$0.10^{+0.07}_{-0.05}$	& 	$0.13_{-0.07}^{+0.09}$	&	$0.31\pm 0.03$	\\
\hline
\end{tabular}
(a) $D = 8\pm 1.4$ kpc \citep{zdz04}; (b) $D=2.39\pm 0.14$ kpc \citep{miller-jones09}; (c) $D=1.7\pm 0.1$ kpc \citep{gelino06}. 
\end{table}
%
%
\section{\so\ in the radio/X-ray domain}

Radio emission from black hole X-ray binaries is generally interpreted as arising from relativistic outflows (\citealt{fender06} and references therein). In the case of extremely low flux density, however, it is worth considering if the radio emission could arise from a background source, or, alternatively, coronal emission from the companion star in the binary system. 

As for A0620-00 \citep{gallo06}, we quantify the probability that the detected radio source is extragalactic in origin using the catalogue of published source counts at 1.4 GHz by \cite{huynh05}. Using a minimum flux density of 10.7 $\mu$Jy, which is the 1.4 GHz flux density corresponding to 4.8 $\mu$Jy at 5.3 GHz for a non-thermal spectrum with index $a=-0.6$, and integrating their equation 12 up to 1000 mJy, yields a probability of $7\times 10^{-3}$ of having an extragalactic background source within the same distance of the predicted position as the detected radio source (0.59\,arcsec). The probability decreases to $8 \times 10^{-4}$ if the distance between the measured X-ray position and the detected radio source is considered (0.2\,arcsec). 

The radio luminosity of a coronally active star depends on its spectral type and rotation period, with generally brighter radio emission for longer rotation periods (this is true for a fixed radius; \citealt{drake89}). The star in \so, which is of spectral type between K5V and K8 \citep{mcclintock01b, wagner01, torres04}, is rotating on the orbital period of 4.08 hr since the orbit is tidally locked. In the catalog of coronally active RS CVn stars of \citet{drake89}, the most luminous system with a rotation period of $\simlt 10$ days, and containing a K5--8 star, would be $1.04 \pm 0.06~\mu$Jy (at 5 GHz) at a distance of $1.7$ kpc. The average radio luminosity of the sample of K5--M1, active RS CVn stars would yield a (5 GHz) flux density of $\sim$$0.4$. We can therefore conclude that the radio emission from XTE J1118+480 is a factor of at least 4 too bright to originate from the corona of the companion, even if the star is coronally active. For the remainder of the paper, we shall assume a jet  origin for the detected radio emission from \so.\\

Next, we compile data from the literature to investigate the behaviour of \so\ in the radio/X-ray luminosity plane over different outbursts, and compare it with that of two well-studied sources; \gx\ and V404 Cyg (see Figure \ref{fig:3sources}).  For \so\ (shown as filled blue circles in Figure \ref{fig:3sources}), we combine hard state observations from the 2000 outburst (data from \citealt{hynes00} and \citealt{fender01}; these were already included in \citealt{gfp03} and GMF12), the new VLA/\cxo\ observations discussed above, plus new data covering the 2005 outburst as reported by \cite{brocksopp10} and \citeauthor{dunn10} (2010; for clarity, the 2005 outburst data are encircled by blue open diamonds in Figure \ref{fig:3sources}), as follows. The observed 4.7 GHz radio flux densities reported by Brocksopp et al. are converted to radio luminosities at 5 GHz (the standard frequency adopted for the compilation of sources) assuming a flat spectrum at all frequencies below 5 GHz, and adopting a distance of 1.7 kpc. X-ray fluxes of \so\ were obtained during the 2005 outburst with the \textit{Rossi X-ray Timing Explorer}. We take the 3-10 keV fluxes reported in \citeauthor{dunn10} and convert them to 1-10 keV fluxes assuming a photon index of $\Gamma = 1.6$ (a typical value for BHBs in the hard state, see, e.g., \citealt{mcclintockremillard}). Lastly, the 1-10 keV fluxes are converted to X-ray luminosities again adopting a distance 1.7 kpc. On eight dates during the outburst decay, radio (4.7 GHz) and X-ray data were taken within 1.0 days of each other.
With the addition of the 2000 outburst data \citep{hynes00,fender01}, and the 2013 quiescent detection reported above, this new data set covers as broad a dynamic range as \gx\ (\citealt{corbel03, corbel13}; open orange diamonds in Figure \ref{fig:3sources}) and V404 Cyg (\citealt{gfp03, corbel08}; open green triangles in Figure \ref{fig:3sources}). 

We then adopt the Bayesian modeling package \textsc{linmix\_err} \citep{kelly07} to investigate the presence of a linear relation of the form (\ellr$-29) = \alpha +\beta$ (\ellx$-35$) for \so, with intrinsic random scatter $\sigma_0$ and radio upper limits included in the fitting. The most likely (hereafter referred to as best-fitting) parameters are estimated as the median of 10,000 draws from the posterior distribution. Quoted 1$\sigma$ errors correspond to the 16$^{\rm th}$ and $84^{\rm th}$ percentiles. 
In order to explore the quantitative influence of random errors (due to e.g. measurement errors, lack of strict simultaneity, model dependent count-rate-to-flux conversions, intrinsic variability, etc.), we run the regression analysis assuming uncorrelated uncertainties on both \ellr\ and \ellx\ of both 0.15 and 0.3 dex (in the case of \so, random errors are likely to dominate over systematic errors caused by distance uncertainties). The best-fitting slope is $\beta=0.72\pm0.04/0.09$, where the quoted error corresponds to the case of 0.15/0.3 uncertainties on the log-luminosities. 
\so\ is thus is the third system -- in addition to \gx\ and V404 Cyg -- for which a strong non linear \lx:\lr\ correlation has been reported over a broad dynamic range. Figure \ref{fig:3sources} also shows a comparison of the the best-fitting slopes estimated by \textsc{linmix\_err} for the three sources (see Table 1 for a complete list of the best-fitting parameters).  

Unlike for \so\ and V404 Cyg, whose distances are established to within 15-20 per cent accuracy (as detailed in \ref{sec:disc}), the distance to GX339-4 is poorly constrained, with $6.7 \simlt D \simlt 9.4$ kpc \citep{zdz04}, corresponding to a factor of up to 2 in luminosity. It is worth noting though that distance uncertainties can not be modelled as uncorrelated errors, as they have the effect of shifting data points for a specific system along the \lx$\propto$\lr\ plane, rather than randomizing the measurements. 
All considered, 0.3 dex, or {a factor $\sim$2 in luminosity seems to be a fair representation of the observed short-term radio and X-ray luminosity variability for hard and quiescent state black hole X-ray binaries} (\citealt{garcia01, gallo06, gallo05, gallo07, hynes03a,hynes04, hynes06, hynes09, bradley07, miller-jones08, bernardini14}; notice that $\sigma \simeq 0.15$ dex was adopted in previous works, e.g. \citealt{gallo12} and \citealt{corbel13}), and it also accounts for the uncertainties in the slope of the partially self-absorbed radio jet (often assumed as flat; please see \ref{sec:disc} for a quantitative discussion). 
%
%
\begin{figure*}
\psfig{figure=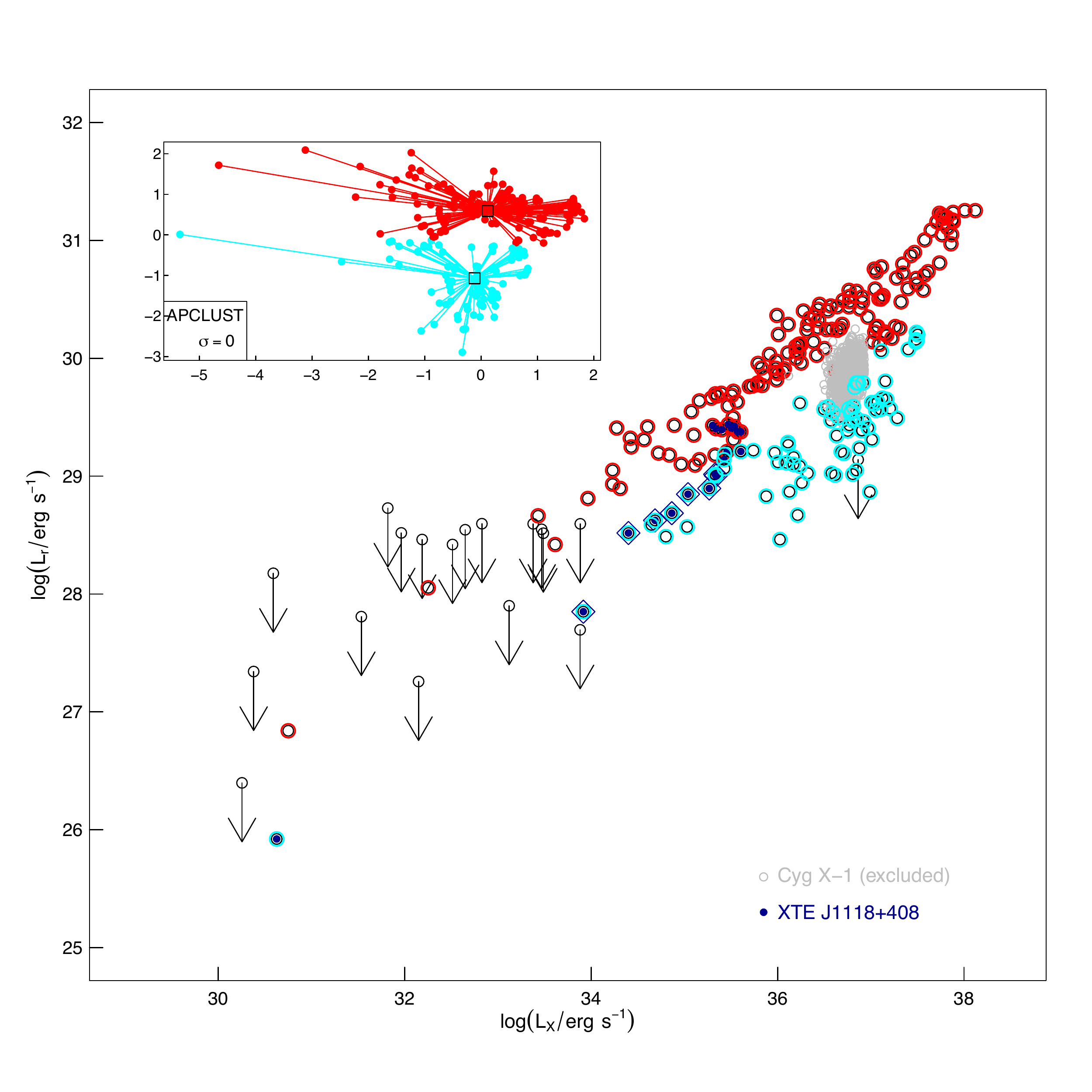,width=0.49\textwidth,angle=0}
\psfig{figure=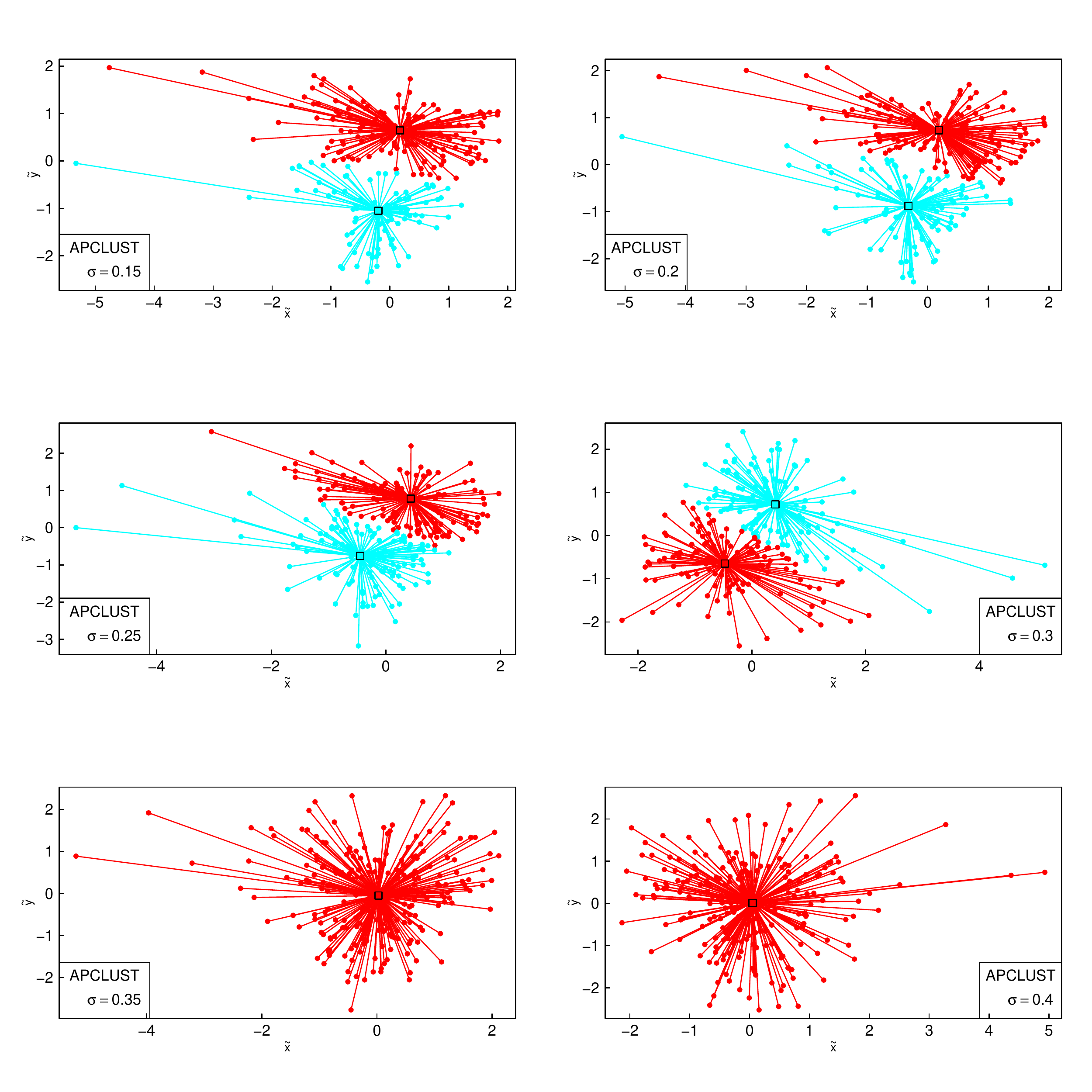,width=0.47\textwidth,angle=0}
\caption{Left panel: Results of the clustering analysis performed with \textsc{APCLUSTER} on the raw data set (i.e, with no random errors applied). The main plot illustrates the results of the standardized-coordinate analysis (shown in the inset) back into the original luminosity plane (in addition to the detections, shown here with open black circles are also upper limits, which are however not included in the clustering analysis). \so\ data are highlighted in dark blue, following the same symbol/colour scheme as in Figure \ref{fig:3sources}. The `cloud' of grey circles around \ellx$\simeq$36.5 and \ellr$\simeq$30, representing the data points for Cyg X-1 in the hard state, is not included in the analysis presented here.
Right panels: results of the clustering analysis after scrambling the data with Gaussian distributions of increasing widths: $\sigma=0.15, 0.2, 0.25, 0.3$ dex. The algorithm ceases to distinguish two clusters for $\sigma=0.35$ dex, corresponding to a luminosity error of a factor $\simgt$2. The colour scheme is defined such that the most populated (`dominant') cluster corresponds to red points; the colour reversal between the $\sigma=0.15, 0.2, 0.25$ panels and the $\sigma=0.3$ panel corresponds to the \textit{bottom} cluster becoming dominant in the latter case.  }\label{fig:apclust}
\end{figure*}
%
%
\section{Clustering and linear regression analysis}
We undertake a formal clustering analysis to characterize different groupings within the radio/X-ray domain of BHBs. In addition to the new data points for \so, the full data set is comprised of the GMF12 plus \citet{corbel13} data, including 24 hard and quiescent state black hole X-ray binary systems, for a total of 247 radio and X-ray detections (the BHB Cyg X-1 is omitted from the analysis at this stage; a separate analysis including Cyg X- is presented in \ref{sec:cygx1}). 
First, the axes are normalised to standardized $x$:$y$ coordinates, as described in Section 3 of GMF12 (see Figure \ref{fig:clust}).  The rotated variables are scaled to unit variance, to ensure comparable dynamic range along both axes. 
We run the clustering algorithms discussed in Section 3.2 of GMF12: `partitioning around medoids' (\textsc{PAM}; \citealt{pam}), `hybrid hierarchical clustering via mutual clusters' (\textsc{hybridHclust}; \citealt{hybridclust}) and `affinity propagation' (\textsc{APCLUSTER}; \citealt{apclust}).  
Taken at face-value, all methods indicate that, qualitatively, a two cluster model provides a better representation of the data. However, as discussed below, \textit{the quantitative results of the clustering analysis are inconclusive}. Firstly, the inferred cluster membership is model-dependent, and thus not robust; for example, shown in Figure \ref{fig:clust} are the groupings identified by \textsc{PAM} (left) and \textsc{hybridHclust} (right). Although both algorithms identify two clusters, they return somewhat different splits about the $y=0$ axis. The reader is referred to Sections 3.1 and 3.2 in GMF12 for a detailed description of the different methods; here we stress that the main difference between \textsc{PAM} and \textsc{hybridHclust} is that the former is a classical partitioning method that separates the data in $k$ clusters, where $k$ is an input  parameter, whereas the latter identifies a top-down clustering diagram with no a priori partition. 
For comparison, the top left inset of the left panel in Figure \ref{fig:apclust} illustrates the groupings as identified by the `affinity propagation' algorithm \textsc{APCLUSTER} \citep{apclust}. The advantage of this method, which is known to outperform the most sophisticated $k$-means clustering algorithms (such as \textsc{PAM}), is that its results do not depend on the initial choice for the subset of points that are first considered as candidate cluster centre (also known as exemplar). The left panel of Figure \ref{fig:apclust} allows to visualize the grouping identified via affinity-propagation in the original \ellr:\ellx\ plane (for clarity, \so's data points are highlighted by blue filled circles, with the 2005 outburst enclosed by blue open diamonds, as in Figure 2). According to this method: $i)$ the full BHB data set is best represented by two clusters; $ii)$ the \so\ data are split into two tracks, with the 2005 outburst decay and the quiescent detection points belonging to the lower track (in cyan); $iii)$ A0620-00, i.e. the only other truly quiescent simultaneous radio/X-ray detection beside \so, is identified instead as belonging to the higher track.  In terms of cluster members, the results of \textsc{APCLUSTER} (Figure \ref{fig:apclust}, top left inset in the left panel) are consistent with those of \textsc{PAM} (Figure \ref{fig:clust}, left panel), but not \textsc{hybridHclust} (Figure \ref{fig:clust}, right panel). 

We have shown that different clustering algorithms applied to this new data set identify somewhat different cluster memberships; this was not the case with the data set analyzed in \citet{gallo12}, where inconsistency between the various cluster memberships were only identified down the clustering tree, i.e. for more aggressive values of the preference value $k$ (in the case of \textsc{APCLUSTER}) or for $k>2$ (in the case of \textsc{PAM}). This means that the addition of new data points has washed out any \textit{robust} 2-cluster partitioning in the radio/X-ray luminosity plane. 
Moreover, we show below that the main result \textit{that the data are best represented by two clusters (regardless of membership) depends critically on the assumptions of the size of random errors}. 
In order to quantify this effect, we randomly scrambled all data points with Gaussian distributions of width $\sigma=0.15, 0.2, 0.25, 0.35$ and 0.4 dex (as discussed in the previous Section, 0.3 dex is likely a better representation of the true random error for this data set than the typically assumed value of $\sigma=0.15$). 
The six square panels on the right hand side of Figure \ref{fig:apclust} show the results of this exercise; firstly, though a two-cluster model is still preferred by the data after scrambling \ellr\ and \ellr\ with a Gaussian distribution of width $\sigma$ up to 0.3 dex, the actual partitioning function is not constant. Most notably, the two cluster exemplars change dramatically between $\sigma=0.25$ and $\sigma=0.3$. Secondly, the bottom cluster becomes more populated than the top cluster for $\sigma=0.3$, whereas the opposite is true for $\sigma$ up to 0.25 (visually, this corresponds to a colour reversal between the $\sigma=0.15, 0.2, 0.25$ panels and the $\sigma=0.3$ panel, as red identifies the largest cluster). Finally, \textsc{APCLUSTER} fails to identify any sub-cluster for $\sigma$ higher than 0.3.  
%
%
\begin{figure*}
\psfig{figure=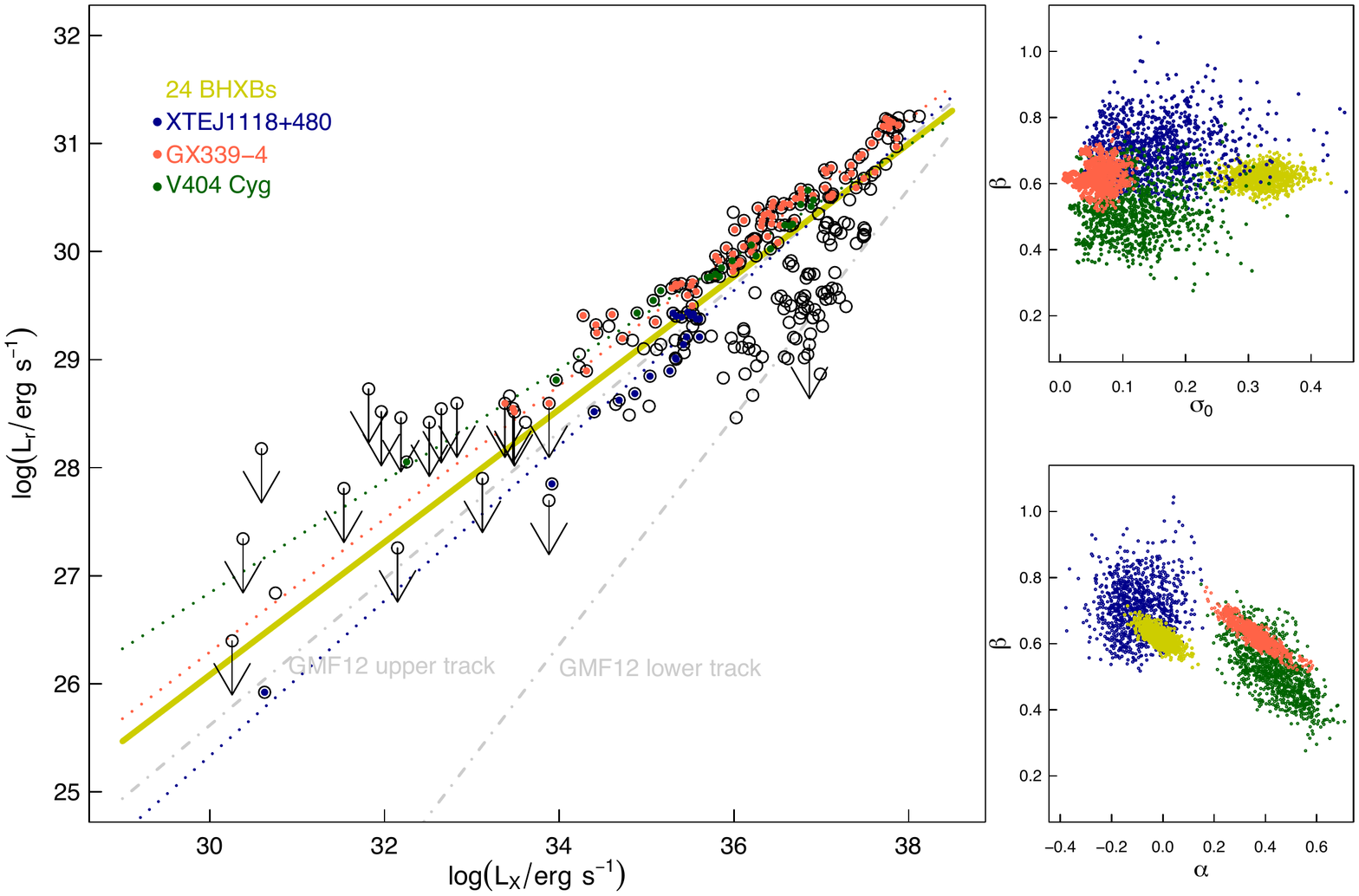,width=\textwidth,angle=0}
\caption{Left panel: Results of the linear regression analysis on the black hole X-ray binaries in the radio/X-ray luminosity domain. A total of 24 systems are included; the thick yellow line represents the best-fitting relation (see Table 1 and text for details on the fitting procedure), to be compared with the upper and lower tracks identified by GMF12 (dashed-dotted gray lines), adopting the same methodology on a smaller data set (18 systems). Also highlighted are the best-fitting relations for V404~Cyg (dotted green line), GX339-4 (dotted orange line) and \so\ (blue dotted line). The two right panels compare the outcomes of the Bayesian fitting routine: shown are 500 draws of the posterior distributions for the slope vs. intrinsic scatter (top) and slope vs. intercept (bottom) for the individual sources and the full sample, according the colour scheme of the left panel. Regardless of how well (or poorly) constrained the slope and intercept are for the individual sources, the best-fitting intrinsic scatter is highest for the full data set, with $\sigma_0 = 0.31\pm0.03$. Quoted best-fitting values are estimated as the median of 10,000 (as opposed to 500) draws, and 1$\sigma$ errors correspond to the 16$^{\rm th}$ and $84^{\rm th}$ percentiles. \label{fig:finale}}
\end{figure*}
In summary, {to the extent that we can rely on clustering algorithms that do not account for upper limits, distance uncertainties and random errors $\simlt 0.3$ dex, a two cluster model offers a better representation of the black hole X-ray binary data set}. Inclusion of random errors of 0.35 dex in both radio and X-ray luminosity has the effect of washing out any statistically significant partitioning for this expanded (with respect to GMF12) dataset. 

Since the clustering analysis results are far from being unique, we abstain from performing regression analysis on given sub-clusters. The validity of a single scaling relation was investigated for the full data set (including upper limits) adopting the same formalism as outlined in the previous Section, with centring on the median luminosities ($\ell_{\rm X,c}=35$ and $\ell_{\rm r,c}=29$).
Figure \ref{fig:finale} shows the results of the Bayesian linear regression analysis on the full data set (in yellow) vs. individual sources (same colours as in Figure 2), also in comparison to the upper and lower track identified by GMF12 (shown as grey dash-dotted lines). The right panels illustrate the posterior distribution of the slope ($\beta$), intercept ($\alpha$) and intrinsic scatter ($\sigma_0$), with the latter ($\sigma_0 = 0.31\pm0.03$) being significantly larger than for the individual sources. 
The best-fitting slope is $\beta=0.61\pm0.03$; see Table 1 for a full list of the best-fitting parameters.
\subsection{Cyg X-1}\label{sec:cygx1}

The BHB Cyg X-1 has been the target of daily, simultaneous radio and X-ray coverage, with the Ryle telescope at 15 GHz and the All Sky Monitor aboard the \textit{Rossi X-ray Timing Explorer}, between 1996 January and 2003 January (see figure 3 in \citealt{gfp03}, and references therein, for details). As a result, this system alone totals over 1000 hard state data points\footnote{As shown in figure 3 of \cite{gfp03}, a softening of the X-ray spectrum is accompanied by the suppression and subsequent quenching of the core radio emission in Cyg X-1; for the purpose of this work, we classified as `hard state' points those corresponding to a (5-12)/(1.5-3) keV  hardness ratio in excess of 1. This is in rough agreement with the more detailed and physically-motivated classification scheme developed by \cite{grinberg13} specifically for Cyg-1.} from daily-averaged fluxes, i.e. more than four times the entire data set discussed above. These are shown as grey circles in Figure \ref{fig:apclust}. While Cyg X-1 can be considered as a peculiar system for a variety of reasons (for example, it comprises a high mass companion, and never fully reaches the soft, thermal dominant state as other sources), none of them is likely to alter the interplay between the radio emission from the compact jet and the X-ray emission from the inner accretion flow in any fundamental way. In other words, there is no good reason for Cyg X-1 to be omitted from our analysis other than practical purposes, in that the number of available data points for this system greatly exceeds that of the other 24 sources combined. 

\begin{figure}
\psfig{figure=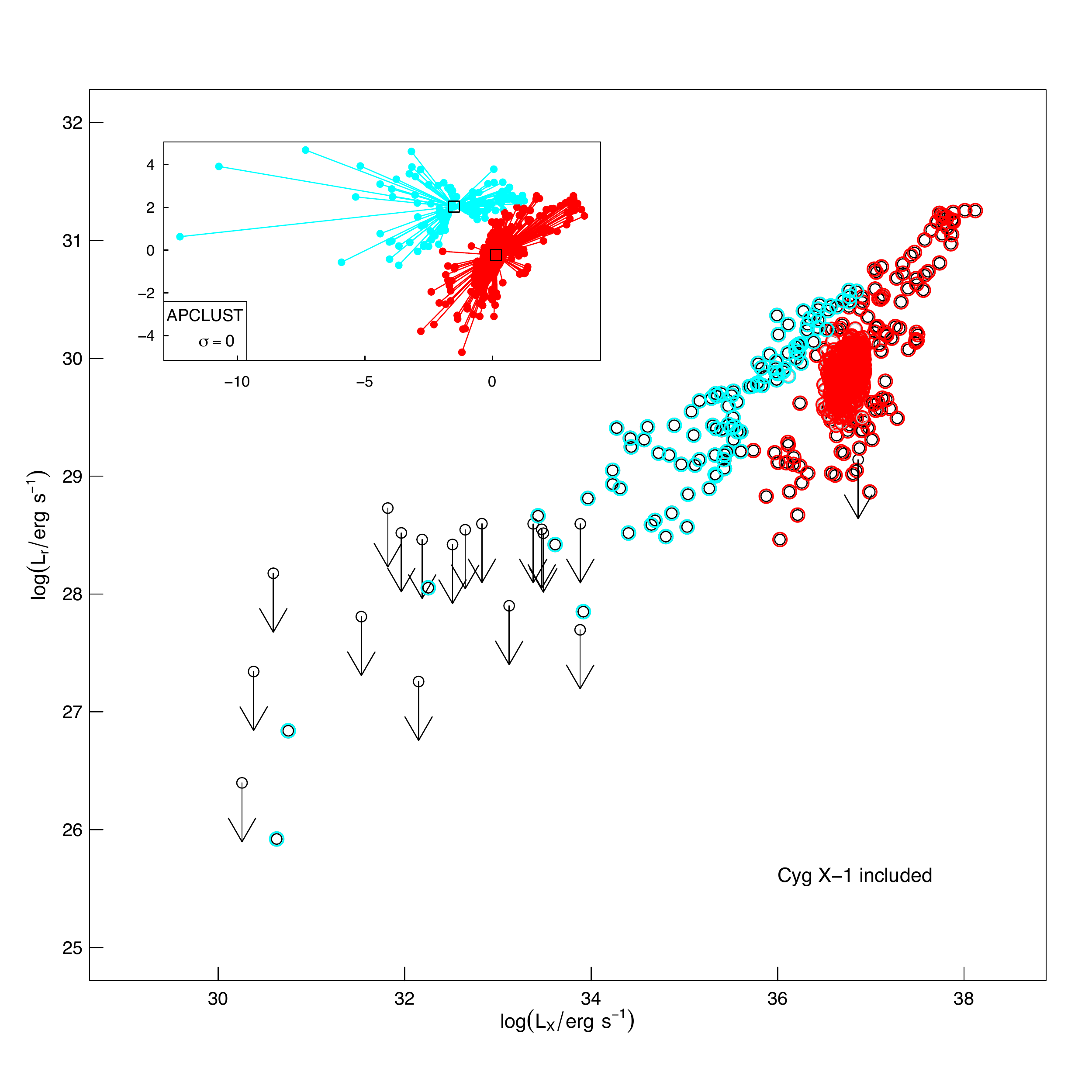,width=0.49\textwidth,angle=0}
\caption{Results from the same clustering analysis as presented in Figure \ref{fig:apclust}, now including additional 1029 data points from the hard state of the high mass X-ray binary Cyg X-1. Although \textsc{APCLUST} still identifies two main clusters within the radio/X-ray luminosity domain, the actual cluster membership appears significantly different from the left panel of Figure 4, where Cyg X-1 was excluded. Most notably, the whole Cyg X-1 data set is part of the bottom cluster (in red), along with a sizable fraction of the data points from the other 24 sources that -- with Cyg X-1 omitted from the analysis -- were instead assigned to the top cluster in Figure 4.}\label{fig:cygx1}
\end{figure}

Admittedly, though, the Cyg X-1 data were excluded from the sample discussed in GFM12. Partially to remedy this omission, here we repeat the clustering analysis for a total of  1276 detections for 25 hard state BHBs, inclusive of 1029 data points from Cyg X-1 (scaled to a distance of 1.86 kpc, from \citealt{reid11}, and corrected for absorption as in \citealt{gfp03}).  The results are shown in Figure \ref{fig:cygx1}; 
not surprisingly, adding so many data points to the initial sample significantly alters the outcome of the clustering analysis. Nevertheless, \textsc{APCLUST} still identifies two main clusters within the standardized coordinate domain. The whole Cyg X-1 hard state data set is characterized as part of the bottom cluster (in red); for the other 24 systems, the main difference in terms of cluster membership with respect to the analysis conducted above (Figure \ref{fig:apclust}) is that a significant fraction of the highest radio and X-ray luminosity data points, corresponding to the top right region of the diagram, is now identified as part of the same cluster as Cyg X-1. Overall, this consolidates the argument for a 2-cluster description of the radio/X-ray domain of hard state BHBs as being somewhat arbitrary, in the sense that the actual cluster membership depends on a number of parameters, where sample selection adds to distance uncertainties, errors and (in some cases) lack of strict simultaneity. 
%
%
%
%
%
\section{Summary and discussion}\label{sec:disc}
Coordinated radio and X-ray monitoring of hard state black hole X-ray binaries has long been advocated as a powerful observational tool for investigating the interplay between radiatively inefficient accretion and the production of steady compact jets in accreting black holes. Over the last decade or so, several groups have collected an impressive amount of data, a large fraction of them being included in this work. 
With the exclusion of Cyg X-1, the sample discussed here is comprised of 24 black hole X-ray binaries, for a total of 265 data points, including 247 detections and 18 upper limits. The most relevant addition compared to previous works is represented by the simultaneous radio/X-ray detection of the nearby, virtually unabsorbed black hole candidate in \so, at about $4\times 10^{-9}$ times its X-ray Eddington luminosity. The quiescent radio counter-part was detected, for the first time, at a level of $4.79\pm1.45~\mu$Jy beam$^{-1}$. At a distance of 1.7 kpc, this corresponds to a monochromatic luminosity of $1.67\times 10^{16}$  \es\ Hz$^{-1}$.  \\

With the addition of \so, a tight (i.e., with inferred intrinsic scatter $\simlt 0.1$) correlation of the form \ellr=$\alpha$$+$$\beta$\ellx\ has now been established for \textit{three} systems, where X-ray luminosities are quoted over the 1-10 keV range, and radio luminosities are integrated up to $\simeq 5$ GHz assuming a flat radio spectrum (see Table 1 and Figure 2). Adopting a Bayesian regression model \citep{kelly07}, and assuming errors of 0.3 dex in both \ellr\ and \ellx, the best-fitting slopes are: $\beta=0.62 \pm 0.04$ for GX339-4 (consistent with the value $0.62\pm0.01$ reported by \citealt{corbel13}); $\beta=0.52 \pm {0.07}$ for V404 Cyg (consistent with the value $0.51\pm0.06$ reported by \citealt{corbel08}), and $\beta=0.72\pm 0.09$ for \so\ (this work; largely based on 2005 data from \citealt{brocksopp10} and \citealt{dunn10}).

Employing a variety of clustering analysis algorithms in order to establish whether a single or multiple cluster model better describes the full data set (24 systems) yields inconclusive results; while two clusters are readily identified by various algorithms, different algorithms identify somewhat different memberships. In addition, a two-cluster description is statistically robust for random errors $\simlt$ 0.3 dex (corresponding to a factor of $\sim$2 in luminosity); when the data are scrambled with a Gaussian distribution of 0.35 dex width, or larger, even the best-performing algorithm starts to return significantly different groupings, or fails to identify multiple clusters (see Figure 4). This is in contrast with the results of GMF12, who identified a robust two-track partition with a data set comprised of 18 black hole X-ray binary systems for a total of 166 data points, indicating that the $\sim$50 per cent increase in the number of data points has washed out any statistically significant partition. 
A linear regression analysis performed on the whole data set returns a best-fitting slope $\beta=0.61\pm0.03$ and intrinsic scatter $\sigma_0=0.31\pm 0.03$. Inclusion of over 1000 data points from long-term, simultaneous radio and X-ray monitoring of Cyg X-1 in the hard state does not affect the above conclusion \textit{qualitatively}.\\

%
%
\begin{figure}
\centering{
\psfig{figure=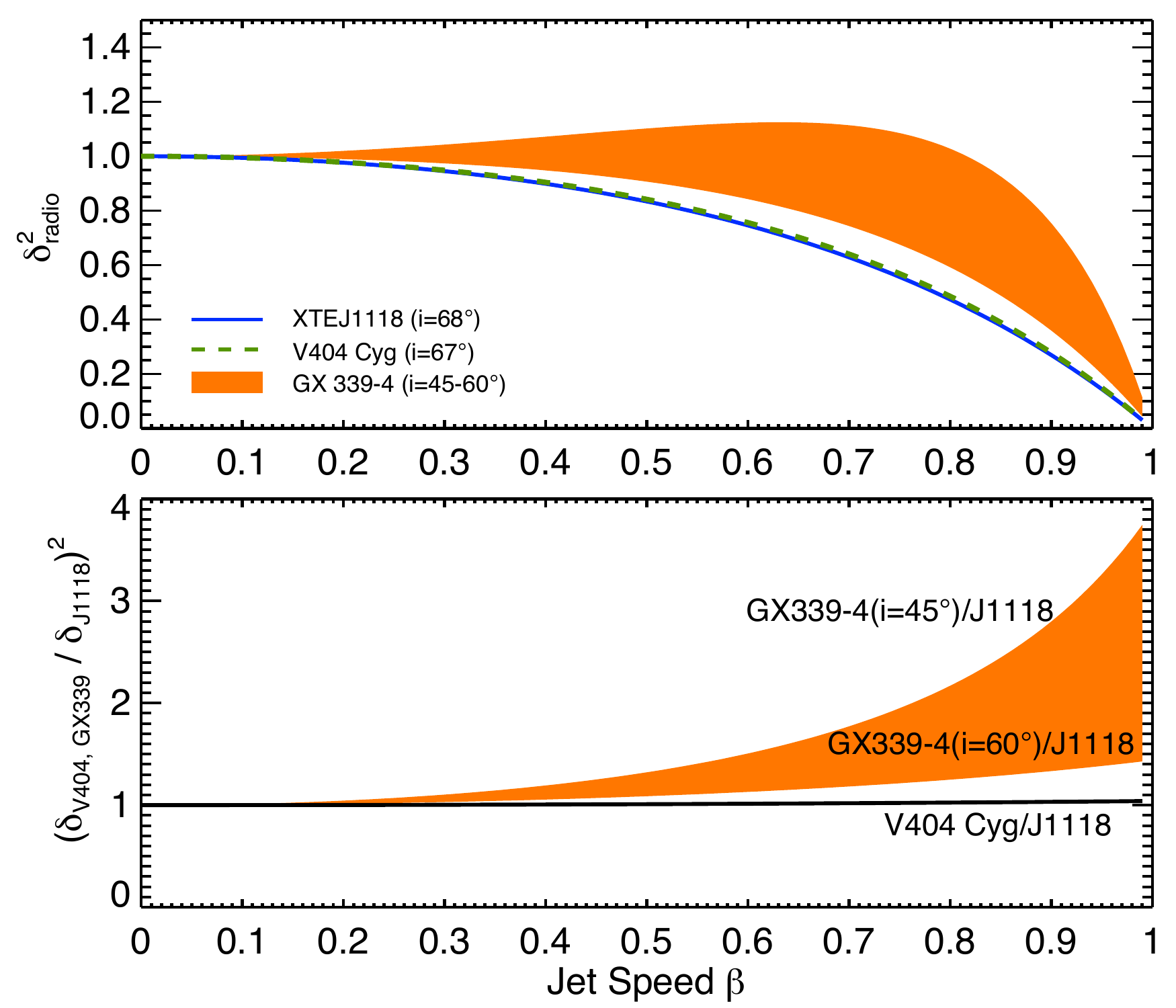,width=0.49\textwidth,angle=0}
\caption{Top: Effective (i.e. averaged between approaching and receding) Doppler boosting factors are shown as a function of the jet intrinsic velocity for \so\ (solid blue line), V404 Cyg (dashed green line) and \gx\ (shaded orange area). Bottom: In attempt to verify whether the difference in the  radio/X-ray correlation normalisation for the three sources may be due to Doppler boosting effects, we plot the ratio of the effective Doppler factors shown in the top panel for V404 Cyg/\so\ (solid line) and \gx/\so\ (shaded area). Doppler boosting alone can not account for the observed difference in normalisation between V404 Cyg and \so, i.e., a factor 0.1 dex, or 25 per cent.  It could however explain the difference between \gx\ and \so\ (i.e., 0.57 dex, or a factor $\sim$4) if the former had a very high intrinsic velocity ($\beta\simgt 0.9$).  \label{fig:boost}}}
\end{figure}

More than ten years after \cite{corbel03} reported on a tight non-linear radio/X-ray correlation in the hard state of GX339-4, and later V404 Cyg \citep{corbel08}, a third system is found to display a similar correlation over a broad dynamic range and over repeated outbursts, thereby strengthening the case for common underlying physics. 
A great deal of work has been done to ascertain the cause of the different normalisations among different systems. In particular,  \cite{solerifender} were able to account for the measured scatter to the radio/X-ray correlation by assuming random jet inclination angles which resulted in highly variable boosting at large Eddington ratios. Here, we adopt a similar approach to investigate whether the difference in normalisation between \so, \gx\ and V404 Cyg may be due to Doppler boosting/de-boosting. In order to carry out a meaningful quantitative comparison, we first re-fit the three data sets by fixing the correlation slope to $0.6$, i.e. the best-fitting value obtained for the whole data set. 
This yields the following values for the best-fitting normalisations: $\alpha_{\rm 0.6}= +0.44, +0.34, -0.13$, for GX339-4, V404 Cyg and \so, respectively (where the centering was fixed at \ellx=35.2 and \ellr=29.2 and typical errors on the quoted slopes are close to 0.15).  Thus, the inferred luminosity ratios range between 0.1 (\gx\ to V404 Cyg) and $\sim$0.6 dex (\gx\ to \so). 
We proceed by assuming that the X-ray emission is un-beamed, and that each system has similar jet properties and velocity profiles at a given X-ray luminosity.  Then, the difference in radio normalisation attributed to Doppler beaming depends only on the orbital inclination of the system (assuming the jet axis is perpendicular to the orbital plane).  Both \so\ and V404 Cyg have well-determined orbital inclinations, of $i=68^{\circ}\pm2^{\circ}$ (\citealt{gelino06}; see also \citealt{khargharia13} for a somewhat broader constraints and \citealt{farr11, kreidberg12} for a comprehensive statistical analysis) and $i={67^{\circ}}^{+3^{\circ}}_{-1^{\circ}}$ \citep{khargharia10, farr11, kreidberg12}, respectively.  The inclination of GX 339-4 is only known to be $\simlt 60$\degree\ \citep{cowley02}, and we assume a range from 45$-$60\degree\ here (see \citealt{zdz04}).  
For a continuously replenished jet, the Doppler-boosted radio luminosity is $L_r = \delta^2 L^\prime_r$, where $L^\prime$ is the un-beamed rest-frame radio luminosity, and $\delta$ is the Doppler factor  (see, e.g., \citealt{mirabelrodriguez}).  The Doppler factor of each system's approaching and receding jet follows $\delta_{\rm rec/app} = \Gamma^{-1}\left(1 \pm \beta \cos \theta\right)^{-1}$, where $\beta$ is the jet speed normalised to the speed of light, $\theta$ is the viewing angle (equal to the orbital inclination), and $\Gamma={1}/{ \sqrt{1-\beta^2}}$ is the bulk Lorentz factor.  Following \cite{gfp03}, owing to the unresolved nature of the radio emission in these systems we estimate the `effective' Doppler factor as $\delta_{\rm radio}^2 = \left(\delta_{\rm app}^2 + \delta_{\rm rec}^2\right)/2$.  This is shown in the top panel of Figure \ref{fig:boost} for each system, as a function of jet speed $\beta$.  In the bottom panel, we show the ratio of Doppler-boosted radio emission from V404 Cyg and GX 339-4 to \so, also as a function of $\beta$, to be compared with the measured normalisation ratios.  The shaded region for GX 339-4 represents the expected range for plausible values due to its uncertain orbital inclination.  The effect of Doppler boosting is {not} sufficiently large to explain the inferred difference in normalisation between V404 Cyg and \so\ (as discussed below, both systems likely have distance errors $\simlt$0.1 dex).  Doppler boosting can only be substantial for GX 339-4 if it has a very fast jet and an orbital inclination toward the lower end of its range.  In that case, however, we would expect to see a difference between the normalisations of GX 339-4 and V404 Cyg, which is not observed. Incidentally, if the X-rays are also beamed, then the difference in beamed radio flux between each system would be even smaller. 

It should be noted that the above analysis, and more generally the quoted uncertainties on the best-fitting intercepts, rely on the quoted values of 1.7$\pm0.1$ and $2.4\pm 0.2$ kpc for \so\ \citep{gelino06} and V404 Cyg \citep{miller-jones09}, respectively. While the latter -- being a parallax distance -- is effectively free of systematic uncertainties, the former uncertainty is likely underestimated; although  \cite{gelino06} assume a K7V type for the donor star in \so\ in their modeling, stellar templates between K5V and K8V are actually consistent with the observed spectrum \citep{torres04}, resulting in a more likely uncertainty of 0.3 kpc (as discussed in \citealt{jonkernelemans}, each spectral class difference translates into a distance uncertainty $\simgt$10 per cent, to be combined with modelling uncertainties). Additionally, the ratios predicted by this method are \textit{curves} rather than parallel lines; in turn, the comparison with the ratios between the best-fitting normalisations can only be taken as indicative of the expected order of magnitude of the effect. \\

Black hole mass and spin are natural and appealing parameters to account for the measured normalisation discrepancies. 
\cite{fender10} found no significant correlation between the correlation normalisations (calculated for all the systems for which data were available at that time, regardless of dynamic range, and assuming a slope of 0.6) and the black hole spin parameters reported in the literature. This conclusion is unaffected by the new \so\ data, as, out of the three systems with well-defined normalisations (i.e. broad dynamic range), \gx\ is the only one with a reported spin parameter value (see \citealt{jmiller04, jmiller08, reis08} and section 5 in \citealt{steiner13} for a comparison between the different spin parameter values inferred via the two leading X-ray spectral fitting methods). If the difference in normalisation between the three systems were driven by black hole spin, then we might expect that the spin parameter of V404 Cyg ought to be comparable to that of \gx, and either should exceed that of \so\ (caveat the larger distance uncertainty for \gx). 
No obvious dependence on black hole mass was identified by \cite{solerifender}; limiting ourselves again to the three systems with broad dynamic range, we note that V404 Cyg and \so\ have remarkably similar -- and relatively narrow -- black hole mass probability distributions\footnote{Notice that, based on the fundamental plane of black hole activity \citep{merloni03, falcke04}, the black hole mass scaling is not exactly linear; a `mass corrected radio luminosity' can be expressed as \lr$/M_{\rm BH}^{0.78}$, with a $\simeq 0.1$ error on the slope.} \citep{farr11, kreidberg12}. Beside the loosely estimated distance, the mass function of \gx\ is also poorly constrained \citep{hynes03b}, making the comparison less interesting at this stage.  \\

Well worth considering is the possibility that systematic differences in the X-ray spectral shape among the different data points might be responsible for rather dramatic excursions across the radio/X-ray domain.  Indeed, not all systems display as regular a behaviour as GX339-4, V404 Cyg and \so.  \cite{coriat11} report on the bizarre evolution of the black hole X-ray binary H1743$-$322 during the decline of its its 2008 outburst  (\citealt{jonker10}, \citealt{miller-jones12}, \citealt{jmiller12b}); during the first phase of the outburst decline, the radio luminosity of H1743--322 appears to decay at a faster pace than in e.g., \gx\ and/or V404 Cyg;  at X-ray luminosities between approximately $10^{36}$ and $10^{34}$ \es, it traces a nearly horizontal excursion in the radio/X-ray plane and reaches a comparable radio luminosity level (for the same \lx) as GX339--4 and V404 Cyg below $\simeq 10^{34}$ \es. 

Based on global scaling relations between the X-ray luminosity and accretion rate in efficient and inefficient accretion flow models \citep{kording06a}, and the radio luminosity and jet power in partially self-absorbed spectrum jet models \citep{heinz03}, this somewhat erratic behavior, and in particular the `radio-quiet' track that H1743-322 describes during the first part of its outburst decline, was interpreted by \cite{coriat11} as due to a phase of radiatively efficient accretion.
Recently, \cite{m-hm14} proposed an explicit form of such an efficient solution (see also \citealt{cao14, huang14}) by arguing that thermal photons from a weak, cool disc in the innermost regions of an otherwise inefficient accretion flow would be responsible for enhancing the seed photons available for Comptonisation, and hence the hard X-ray flux (in the context of this model, the `radio quiet' track corresponding to the initial decline of the 2008 outburst of H1743-322 is better described as `X-ray bright').   The condensation of hot, optically thin accreting gas into an inner, cool, keplerian disc is predicted above a critical mass accretion rate if thermal conduction and Compton cooling are properly accounted for in the equations of energy and mass exchange \citep{meyer07, liu06, liu07}. 
From a theoretical standpoint thus, no inner disc is expected at low accretion rates, where (for reasonable values of the viscosity parameter) the threshold can be set around $10^{-3}$ times the Eddington limit. 

Although this is exactly what was observed in the case of H1743-322, the new observations of \so\ and other systems presented here make this appealing interpretation less clear, in the sense that the neat distinction between two tracks claimed by GMF12 is no longer robust (though one could argue that \so, too, experienced a brief period of radiatively efficient accretion, corresponding to the 2000 outburst points, i.e., the blue filled circles that are {not} enclosed by open diamonds in Figure 4). Also, it is important to note that observational support to the existence of cool inner discs in hard state black hole X-ray binaries has been claimed for a several systems \citep{disalvo01,jmiller06a, reis09, reis10}, albeit \cite{tomsick09} have shown that the inner disc in \gx\ recedes sharply below 1 per cent of the Eddington luminosity, i.e. over the luminosity range where this system exhibits a remarkably tight non-linear correlation (\citealt{corbel13}; however, see, e.g., \citealt{dangelo08} and \citealt{kolehmainen14} for a different interpretation of the soft X-ray excess). 

To add to the above considerations, a word of caution is in order about the implicit assumption of flat radio spectrum that is typically folded into the radio luminosity measurements quoted in this and other works. While remarkably flat radio spectra have been measured for a handful of sources, and are known to persist over several orders of magnitude in luminosity (e.g., \citealt{fender00, gallo05, russellbreaks}), hard state BHBs are known to exhibit \textit{flat-to-inverted}\footnote{Here defined as having a positive slope, $a$, where the monochromatic flux density $S_{\nu}$ scales as $\nu^{+a}$.} radio spectra \citep{fender00}. Spectral indices as inverted as $a=+0.5$ have been reported just for \so~during various phases of its 2000 outburst (\citealt{hynes00, fender01}; see also \citealt{russellt14} and \citealt{corbel13}, reporting on inverted radio spectra for MAXI J1836--194 and \gx, respectively, while in the hard state).  Even assuming (for practical purposes) that the radio spectra of hard and quiescent black hole X-ray binaries extend from a minimum frequency that is much lower than the observing frequency for all sources  ($\nu_{\rm min}<<\nu_{\rm obs}$), adopting a flat (i.e., $a=0$) vs. inverted (e.g., $a=+0.5$) spectral index introduces a factor of 2 upward error in the estimate of the integrated luminosity (which, in the case of flat spectrum, is simply calculated by multiplying the measured monochromatic flux density by the observing frequency).

Adding to this is another -- potentially more problematic -- issue. The origin of the X-ray emission in quiescence may differ from the hard state; In a companion paper (Plotkin \etal, submitted), we combine the X-ray and radio observations presented in this work with simultaneous near-infrared and ultraviolet observations (with \textit{Swift}) and (non-simultaneous) IR observations (with \textit{Spitzer}). When fitted with a multi-zone jet model \citep{markoff01,markoff03,markoff05}, the SED is consistent with a Synchrotron-Self-Compton (SSC) origin for the X-ray power law. In contrast, modeling the broadband SED of \so\ at \ratio$\simeq 10^{-5}$ with the same model favours a predominantly synchrotron origin for the X-ray emission \citep{maitra09}. 
Our results indicates that, as the system's bolometric luminosity decreases towards quiescence, the jet becomes progressively less magnetically dominated, and the accelerated non-thermal electron tail reaches lower Lorentz factors, i.e., the particle acceleration process becomes less efficient. \\

In closing, the new \so\ data, including the first simultaneous radio/X-ray detection at \lx/\ledd$\simeq 10^{-8.5}$, have confirmed the existence of a strong non-linear radio/X-ray luminosity correlation for \textit{individual} hard and quiescent state sources. At the same time, new data have weakened the evidence for two well-defined tracks in the radio/X-ray domain of BHBs. It is entirely possible the data set represented in Figure \ref{fig:finale}, albeit indicative of an overall trend of radio luminosity increasing non-linearly with X-ray luminosity for sub-Eddington systems, encloses several, potentially uncorrelated phenomena that may be at the origin of its large scatter. We conclude by noticing that multiple, coordinated radio/X-ray observations of two super-massive black holes in nearby Seyfert galaxies and low-luminosity AGN, performed over periods of several months \citep{bell11, king11, king13}, show how these systems (both akin the most radio-quiet BHBs in terms of radio-to-X-ray luminosity ratios) make almost orthogonal excursions across the fundamental plane best-fitting relation from \cite{gultekin09}, further strengthening the argument against (uncertainties in the) black hole mass or spin parameter as being entirely responsible for the large  inferred intrinsic scatter. 

%
\section{Acknowledgements}
Support for this work was provided by the National Aeronautics and Space Administration through Chandra Award Number GO3-14036X issued by the Chandra X-ray Observatory Center, which is operated by the Smithsonian Astrophysical Observatory for and on behalf of the National Aeronautics Space Administration under contract NAS8-03060. JCAMJ acknowledges support from an Australian Research Council Discovery Grant (DP120102393). SC acknowledges funding support from the French Research National Agency: CHAOS project ANR-12-BS05-0009 and financial support from the UnivEarthS Labex program of Sorbonne Paris Cit\'e (ANR-10-LABX-0023 and ANR-11-IDEX-0005-02). The National Radio Astronomy Observatory is a facility of the National Science Foundation operated under cooperative agreement by Associated Universities, Inc. 
%
%
\bibliographystyle{mn2e}


\end{document}